\documentclass[conference]{IEEEtran}
\IEEEoverridecommandlockouts
\usepackage{cite}
\usepackage{amsmath,amssymb,amsfonts}
\usepackage{algorithmic}
\usepackage{graphicx}
\usepackage{textcomp}
\usepackage{xcolor, soul}
\sethlcolor{yellow}
\usepackage{comment}

\usepackage{siunitx}
\usepackage{hyperref}
\usepackage[nolist]{acronym}
\usepackage[flushleft]{threeparttable}
\usepackage{placeins}
\usepackage{svg}  
\usepackage{float}
\usepackage[percent]{overpic}
\usepackage{subcaption}

\usepackage{colortbl}
\definecolor{Gray}{gray}{0.85}

\usepackage{mathtools}
\usepackage{multirow}
\usepackage{booktabs}
\usepackage[inline]{enumitem}

\usepackage{booktabs,tabularx,siunitx}
\usepackage{caption}
\usepackage[font=footnotesize,labelfont=bf]{caption}
\usepackage{orcidlink}  
\usepackage{textcomp}   
\usepackage{eso-pic}    

\newcommand{\revise}[1]{{\textcolor{black}{#1}}}     

\newcommand{\RNum}[1]{\uppercase\expandafter{\romannumeral #1\relax}}  

\DeclareRobustCommand{\IEEEauthorrefmark}[1]{\smash{\textsuperscript{\footnotesize #1}}}

\DeclareSIUnit\dbi{dBi}                            
\DeclareSIUnit\dbm{dBm}                            
\DeclareSIUnit\msInference{ms/inference}           
\DeclareSIUnit{\MBPS}{MBPS}                        
\DeclareSIUnit{\MSPS}{MSPS}                        
\DeclareSIUnit{\Div}{Div}                          

\def\BibTeX{{\rm B\kern-.05em{\sc i\kern-.025em b}\kern-.08em
    T\kern-.1667em\lower.7ex\hbox{E}\kern-.125emX}}
\begin{document}

\bstctlcite{IEEEexample:BSTcontrol}

\AddToShipoutPictureBG*{
  \AtPageUpperLeft{%
    \put(0,-40){\raisebox{15pt}{\makebox[\paperwidth]{\begin{minipage}{21cm}\centering
      \textcolor{gray}{This article has been accepted for publication in the proceedings of the \\
      International Conference on Body Sensor Networks (IEEE-EMBS BSN 2025).\\ 
      } 
    \end{minipage}}}}%
  }
  \AtPageLowerLeft{%
    \raisebox{25pt}{\makebox[\paperwidth]{\begin{minipage}{21cm}\centering
      \textcolor{gray}{ \copyright 2025  Authors and IEEE. 
       This is the author’s version of the work. It is posted here for your personal use. Not for redistribution. The definitive Version of Record will
        be published in the Proceedings of the International Conference on Body Sensor Networks (IEEE-EMBS BSN 2025).
      }
    \end{minipage}}}%
  }
}

\title{Wireless Low-Latency Synchronization for Body-Worn Multi-Node Systems in Sports}


\author{
\IEEEauthorblockN{
Nico Krull\IEEEauthorrefmark{1},
Lukas Schulthess\IEEEauthorrefmark{1}$^*$\orcidlink{0000-0002-6027-2927},
Michele Magno\IEEEauthorrefmark{1}\orcidlink{0000-0003-0368-8923},
Luca Benini\IEEEauthorrefmark{1}\IEEEauthorrefmark{2}\orcidlink{0000-0001-8068-3806},
Christoph Leitner\IEEEauthorrefmark{1}$^*$\orcidlink{0000-0002-7058-7236}\thanks{$^*$Corresponding authors: L. Schulthess (lukas.schulthess@pbl.ee.ethz.ch), C. Leitner (e-mail: christoph.leitner@iis.ee.ethz.ch).}
}
\IEEEauthorblockA{
\textit{\IEEEauthorrefmark{1}Department of Information Technology and Electrical Engineering, ETH Zurich, Switzerland}\\
\textit{\IEEEauthorrefmark{2}Department of Electrical, Electronic, and Information Engineering, University of Bologna, Italy}
}
}

\maketitle


\begin{abstract}
Biomechanical data acquisition in sports demands sub-millisecond synchronization across distributed body-worn sensor nodes. \revise{This study evaluates and characterizes the Enhanced ShockBurst (ESB) protocol from Nordic Semiconductor under controlled laboratory conditions for wireless, low-latency command broadcasting, enabling fast event updates in multi-node systems.}
Through systematic profiling of protocol parameters, including cyclic-redundancy-check modes, bitrate, transmission modes, and payload handling, we achieve a mean \ac{D2D} latency of 504.99 \textpm{} 96.89 \textmu{}s and a network-to-network core latency 
of 311.78 \textpm{} 96.90 \textmu{}s using a one-byte payload with retransmission optimization.
This performance significantly outperforms Bluetooth Low Energy (BLE), which is constrained by a \SI{7.5}{ms} connection interval, by providing deterministic, sub-millisecond synchronization suitable for high-frequency (\SIrange{500}{1000}{\hertz}) biosignals.
These results position ESB as a viable solution for time-critical, multi-node wearable systems in sports, enabling precise event alignment and reliable high-speed data fusion for advanced athlete monitoring and feedback applications.
\end{abstract}

\vspace{4pt}
\begin{IEEEkeywords}
Sport, Biomechanics, WBAN, Wireless, Wearable, Sensor
\end{IEEEkeywords}

\begin{acronym}
    \acro{RF}{Radio Frequency}
    \acro{IoT}{Internet of Things}
    \acro{IoUT}{Internet of Underwater Things}
    \acro{UWN}{Underwater Wireless Network}
    \acro{UWSN}{Underwater Wireless Sensor Node}
    \acro{AUV}{Autonomous Underwater Vehicles}
    \acro{UAC}{Underwater Acoustic Channel}
    \acro{FSK}{Frequency Shift Keying}
    \acro{OOK}{On-Off Keying}
    \acro{ASK}{Amplitude Shift Keying}
    \acro{UUID}{Universal Unique Identifier}
    \acro{PZT}{Lead Zirconium Titanate}
    \acro{AC}{Alternating Current}
    \acro{NVC}{Negative Voltage Converter}
    \acro{NVCR}{Negative Voltage Converter Rectifier}
    \acro{FWR}{Full-Wave Rectifier}
    \acro{MCU}{Microcontrollers}
    \acro{GPIO}{General Purpose Input/Output}
    \acro{PCB}{Printed Circuit Board}
    \acro{AUV}{Autonomous Underwater Vehicle}
    \acro{IMU}{Intertial Measurement Unit}
    \acro{BLE}{Bluetooth Low Energy}
    \acro{FSR}{Force Sensing Resistor}
    \acro{SiP}{System in Package}
    \acro{SoC}{System on Chip}
    \acro{SpO2}{Oxigen Saturation}
    \acro{PULP}{Parallel Ultra-Low Power}
    \acro{ML}{Machine Learning}
    \acro{ADC}{Analog to Digital Converter}
    \acro{TCDM}{Tightly Coupled Data Memory}
    \acro{ESB}{Enhanced ShockBurst}
    \acro{UART}{Universal Asynchronous Receiver/Transmitter}
    \acro{IPC}{Inter-Process Communication}
    \acro{APP-core}{Application Core}
    \acro{NET-core}{Network Core}
    \acro{ACK}{Acknowledgment}
    \acro{JSON}{JavaScript Object Notation}
    \acro{CRC}{Cyclic Redundancy Check}
    \acro{FIFO}{First-In, First-Out}
    \acro{RX}{receiver}
    \acro{TX}{transmitter}
    \acro{Rx}{receive}
    \acro{Tx}{transmit}
    \acro{D}{Digital input channel of the oscilloscope}
    \acro{SD}{Standard Deviation}
    \acro{OLCfg}{Optimal Latency Configuration}
    \acro{IMU}{Inertial Measurment Unit}
    \acro{EMG}{Electromyography}
    \acro{GNSS}{Global Navigation and Satelite System}
    \acro{WPAN}{Wireless Personal Area Network}
    \acro{WBAN}{Wireless Body Area Networks}
    \acro{ESB}{Enhanced ShockBurst}
    \acro{IPC}{Inter-Process Communication}
    \acro{SCPI}{Standard Commands for Programmable Instruments}
    \acro{D2D}{Device-to-Device}
    \acro{ISM}{Industrial, Scientific, and Medical}

\end{acronym}
\section{Introduction}\label{sec:introduction}

Human movements, particularly in sports, involve rapid and complex motion sequences \cite{j_pastel_complex_movement_2022}. Relying solely on single-source data (i.e., cameras) is often insufficient for assessing the quality of movements \cite{j_swain_multi_sensor_feedback_2023}. Consequently, multi-sensor-multi-node systems have become essential tools for analysis in sports applications \cite{j:yang2024}. These systems measure critical biomechanical variables, such as segment accelerations, forces, and muscle activation, in real-time, utilizing samples collected from unobtrusive, distributed sensor nodes on the athlete's body or embedded into sporting equipment. By gathering this information from sensors such as e.g. \acp{IMU}, pressure insoles \cite{c:schulthess2023}, or \ac{EMG} devices \cite{c:wu2023}, a more comprehensive understanding of an athlete's movement and performance can be obtained. However, accurate data fusion across multiple nodes demands tight synchronization, particularly because biomechanical data acquisition often requires sampling rates ranging from several hundred \qty{}{\hertz} (e.g., \ac{IMU}, pressure sensors) to over 1 \qty{}{\kilo\hertz} (e.g., \ac{EMG}) \cite{j:weber2025}. Consequently, synchronization errors below \qty{1}{\milli\second} are key. This ensures that high-speed biomechanical events are accurately positioned in time across all sensor data streams, enabling precise temporal correlation and on-device processing or post-action analyses. 

\ac{BLE} is a widely adopted communication standard in \acp{WBAN} due to its wide availability and easy integration. With a low power consumption in the \qty{}{\milli\watt} range, and a relatively high throughput of up to \qty{2}{\mega\bit\per\second}, and the ability to support multi-node connectivity, it is well-suited for communication among wearables \cite{j_kim_wireless_2025}. However, \ac{BLE}'s communication scheduling, based on coarse-grained connection intervals \cite{Bulic2017} and adaptive frequency hopping, inherently introduces latency and jitter \cite{w:BLE1}. Consequently, achieving deterministic sub-millisecond synchronisation with unmodified \ac{BLE} remains challenging \cite{c:ohara_wireless_timesync_2024}. While connection intervals for packet transmission can be set statically, they are limited to a lower bound of \qty{7.5}{\milli\second} \cite{w:BLE2}, which restricts fast, short-term event triggering. To address these limitations, previous works have explored analogue current profiling of \ac{TX} and \ac{RX} as an indirect method for temporal alignment via crystal drift correction  \cite{j:weber2025}. However, implementing this approach in practice requires additional hardware-assisted workarounds outside the \ac{BLE} stack. In contrast, the \ac{ESB} protocol is designed for low-latency, low-power, short-range wireless communication, offering a promising alternative for fast data transmission on the \qty{2.4}{\giga\hertz} \ac{ISM} band, enabling sub-millisecond latency for time-critical applications \cite{nordic_esb}.

The main contribution of this study is an accurate performance profiling of the \ac{ESB} protocol in a controlled laboratory setting using two \textsc{Nordic Semiconductor nRF5340} development boards. By probing signals at key points within the software stack of the dual-core architecture of both \ac{TX} and \ac{RX} nodes using \ac{GPIO} signaling, we identify optimal configurations for achieving deterministic, low-latency synchronisation suitable for high-frequency biomechanical applications.

\begin{figure*}[t]
    \centering
    \begin{overpic}
        [width=\textwidth]{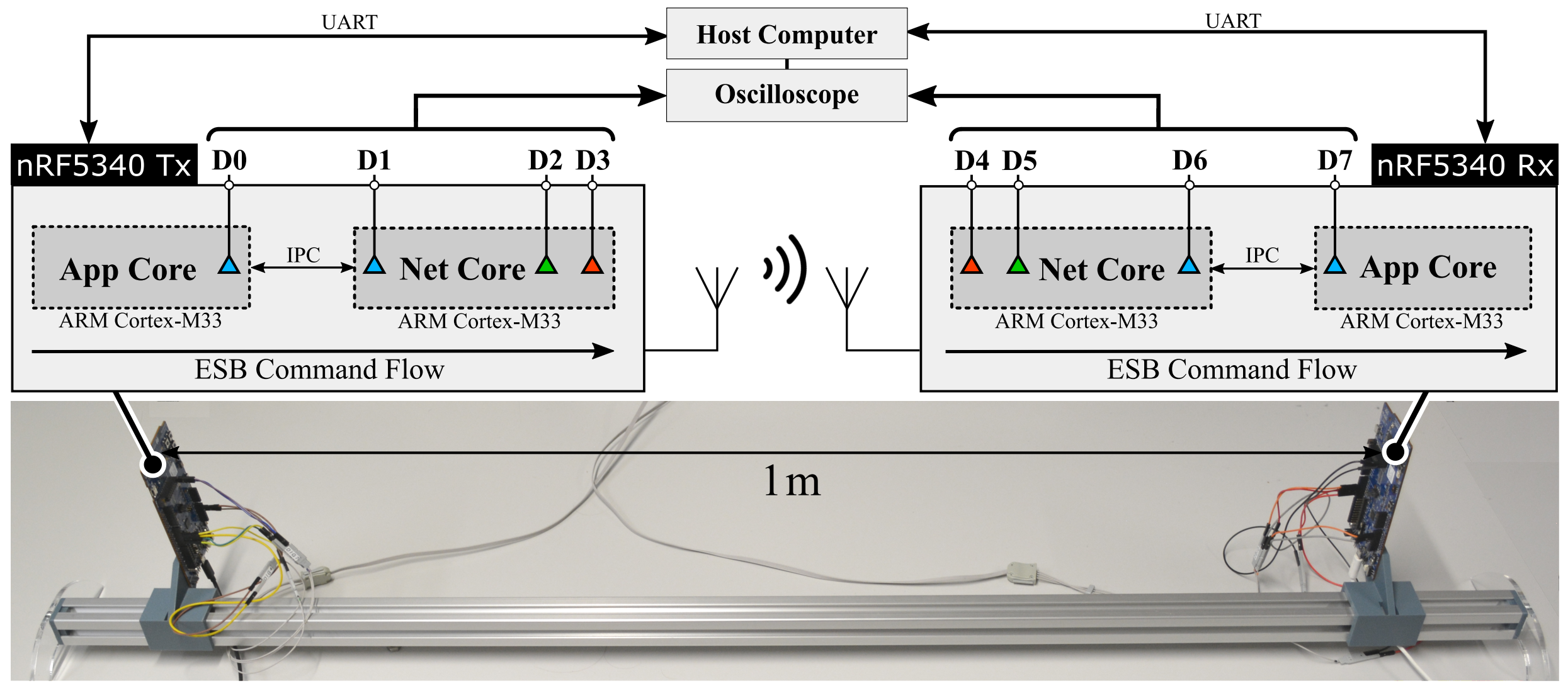}
    \end{overpic}
    {\footnotesize
        \begin{tabular}{ll@{\hskip 0.5cm}ll}
    \multicolumn{2}{c}{\textbf{Transmitter}} & \multicolumn{2}{c}{\textbf{Receiver}} \\
        \textbf{D0:} & IPC transmit initiated (app core) & 
        \textbf{D4:} & First command in ESB library after reception (net core) \\
        \textbf{D1:} & IPC received (net core) & 
        \textbf{D5:} & ESB event handler received payload (net core) \\
        \textbf{D2:} & After ESB command, send payload (net core) & 
        \textbf{D6:} & IPC transmit initiated (net core) \\
        \textbf{D3:} & Last command in ESB library before transmission (net core) & 
        \textbf{D7:} & IPC received (app core)
\end{tabular}
    }
    \caption{Overview of the measurement setup and the \ac{GPIO} trigger point locations.}
    \vspace{-4mm}
    \label{fig:setupoverview}
    \vspace{-2mm}
\end{figure*}

\section{Methods}\label{sec:methods}
The \ac{ESB} protocol \cite{nordic_esb} is a low-power, low-latency wireless communication protocol from Nordic Semiconductor offering bidirectional data transfer with packet queuing in \ac{TX} and \ac{RX} \ac{FIFO} buffers, acknowledgements, and automatic retransmissions of lost packets. 

For the characterization of \ac{ESB} parameters, the payload size was set to 8 bytes of bit-encoded command data. Moreover, acknowledgement has been disabled to allow broadcast communication without waiting for responses. \ac{TX} and \ac{RX} devices are configured on the same \ac{ESB} data pipe for uniform reception. The retransmission parameter was set to two, meaning that each packet is transmitted a total of three times, with a minimum delay of \qty{435}{\micro\second} between consecutive transmissions.

\subsection{Parameter Configuration}\label{sec:parameters}
The following \ac{ESB} protocol parameters were investigated for their impact on latency performance:

\begin{description}
\item[CRC mode:] \ac{CRC} can be configured to 16 bits, 8 bits, or disabled entirely, affecting error detection reliability and effective payload size.
\item[Protocol mode:] can be configured as either fixed or dynamic payload length.
\item[Bitrate mode:] can be configured to \qty{1}{\mega\bit\per\second} or \qty{2}{\mega\bit\per\second}, either using standard settings or \ac{BLE}-specific modes with adjusted radio parameters.
\item[\ac{TX} mode:] offering automatic, manual, and manual start modes, influencing how packets are dequeued from the \ac{TX} \ac{FIFO}.
\item[\ac{TX} power levels:] 
adjustable from \qty{-70}{\dbm} and \qty{10}{\dbm}.
\end{description}

Besides optimizing the \ac{ESB} protocol parameters, the payload construction was also refined to further reduce latency. In the standard setup, the \ac{APP-core} generates the data and transfers it to the \ac{NET-core} via \ac{IPC} before it is written to the \ac{TX} \ac{FIFO}. In the mode optimised for time-critical execution, a minimal payload size of 1-byte is pre-constructed directly on the \ac{NET-core}, allowing it to remain ready for immediate transmission upon command. This approach eliminates the need for additional \ac{IPC} transfers and runtime payload formatting, thereby reducing transmission latency. For a fair comparison between the standard and optimized payload construction modes, the payload size was fixed at 1-byte for these investigations.

\subsection{Measurement Setup and Characterization}
The measurement setup consisted of four key components: (1) a digital oscilloscope (MSOX3024T, Keysight Technologies Inc.), two \textsc{nRF5340DK} development boards (Nordic Semiconductor) as \ac{TX} (2) and \ac{RX} (3), and a host computer (4). \revise{Both \ac{TX} and \ac{RX} boards were mounted on a custom linear rail, spaced \qty{1}{\meter} apart with upward-facing, opposing antennas to ensure consistent and comparable measurements (see \autoref{fig:setupoverview}).} A host computer orchestrated the test setup via a Python script. The computer handled communication with the \ac{TX} and \ac{RX} boards by sending configuration commands via \ac{UART} and retrieving measurements from the oscilloscope for post-analysis using \ac{SCPI}. 
The \textsc{nRF5340} \ac{MCU} features an \ac{APP-core} and a \ac{NET-core}, with the \ac{ESB} application running entirely on the \ac{NET-core} and requiring data transfer between cores via \ac{IPC}. To capture precise event timing, \ac{GPIO} pins were toggled at key stages of the Tx-Rx process and recorded by the oscilloscope (referred to as \ac{D}, see caption of Figure~\ref{fig:setupoverview}). Meanwhile, the \ac{RX} board also returned the collected payload to the host computer for further analysis.

\subsection{Measurement Protocol and Analyses}
Each parameter in Section~\ref{sec:parameters} was independently evaluated for its impact on transmit latency, with all other parameters held constant. All possible configuration options were tested, and their order was randomly shuffled before each test round to minimize systematic bias.
Depending on the number of configurations, three to five rounds were conducted, each recording 150 \ac{TX} attempts to ensure statistical reliability.
The oscilloscope was configured to trigger on the rising edge of D3, with a time base of \qty{600}{\micro\second/\Div} for a \qty{6}{\milli\second} capture window and 60000 sampled points per transmission, providing a temporal resolution of \qty{\pm 0.1}{\micro\second}. This measurement setup was kept consistent across all tests to ensure high precision in latency evaluation.
For the experimental data analysis, the mean, median, and \ac{SD} were calculated.
\section{Results and Discussion}\label{sec:results}
To determine the \ac{OLCfg} concerning latency, various \ac{ESB} parameters were evaluated and compared.

\begin{figure}[hb]
    \centering
    \includegraphics[width=1\linewidth]{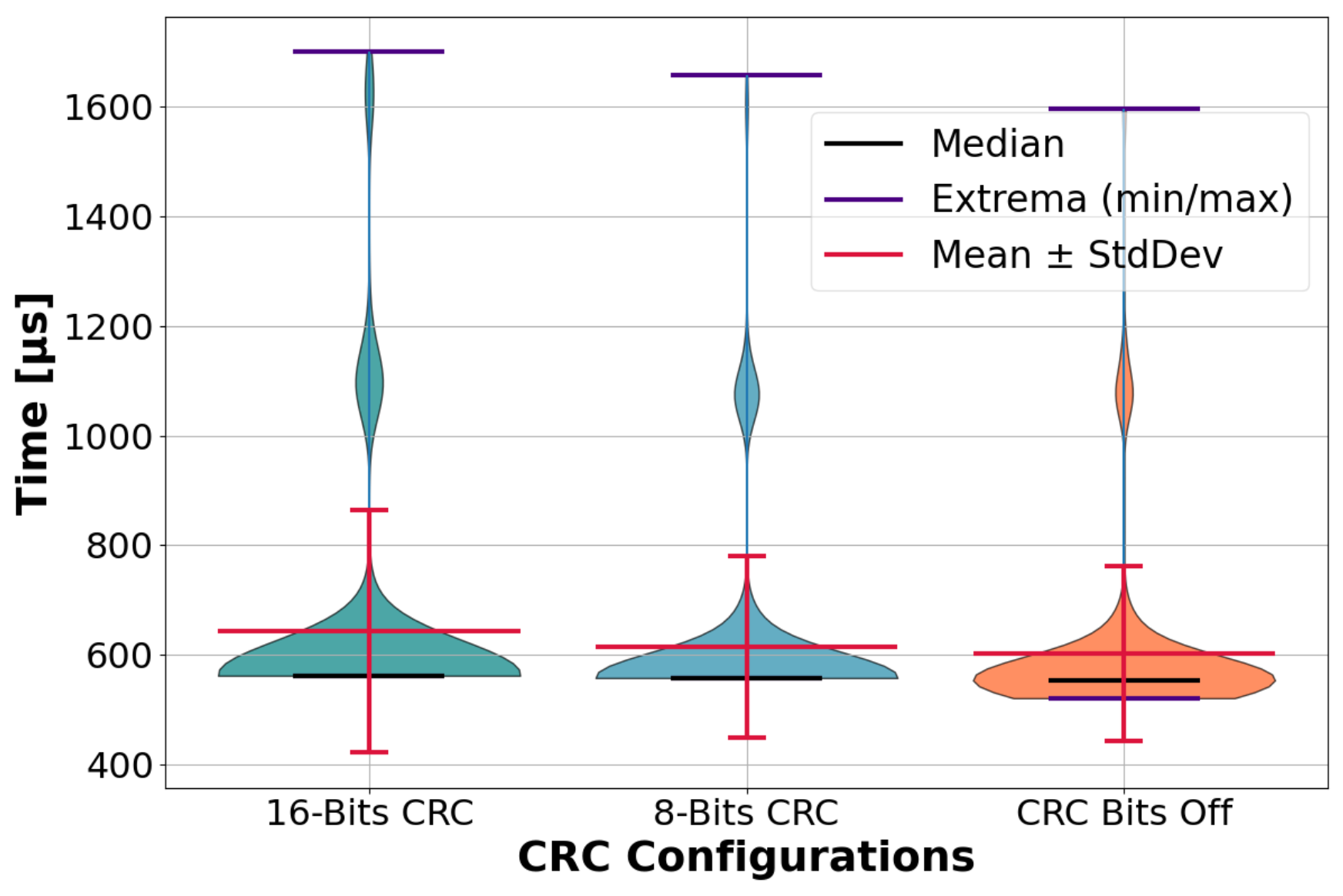}
    \caption{Distribution of the time difference between the digital inputs D0 and D7 for the three \ac{CRC} configurations \SI{8}{\bit},\SI{16}{\bit}, disabled (8-byte payload, 750 transmissions per configuration).}
    \label{fig:crc_time_plot}    
\end{figure}
\vspace{-5mm}

\begin{table}
\centering
\caption{Number of packets transmitted and received under the three CRC configurations. An 8-byte payload was sent in five rounds of 150 transmissions per configuration.}
\label{tab:crc_amount}
\begin{tabularx}{\columnwidth}{l *{3}{>{\centering\arraybackslash}X}}
\toprule
& \textbf{16-bit CRC} & \textbf{8-bit CRC} & \textbf{CRC off} \\
\midrule
\textbf{Sent}     & 750 & 750 & 750 \\
\textbf{Received} & 736 & 740 & 735 \\
\textbf{Unique}   & 736 & 740 & 734 \\
\textbf{Valid}    & 736 & 740 & 719 \\
\bottomrule

\vspace{-8mm}

\end{tabularx}
\end{table}

\subsection{\ac{ESB} Parameter Analyses}

\subsubsection{\acl{CRC} Analysis}
The \ac{CRC} mode directly influences payload length and checksum computation time, thereby affecting overall transmission latency. Three \ac{CRC} configurations are supported: \SI{16}{\bit}, \SI{8}{\bit}, and \ac{CRC} disabled entierly. 
Figure~\ref{fig:crc_time_plot} shows the total transmission time, measured as the time difference between \ac{D}0 and \ac{D}7. It defines the duration from issuing the transmission command on the \ac{APP-core} of the \ac{TX} to receiving the reception notification on the \ac{APP-core} at the \ac{RX}. All measurements were conducted with two retransmissions configured, generating three bulges in the violin plot. The first bulge, occurring at around \SI{600}{\micro\second}, corresponds to packets from the first transmission, where the upper two originate from retransmissions one and two.

Table~\ref{tab:crc_amount} shows that the overall number of transmitted and received packages remains relatively stable across all modes, with variations below 0.7\% likely attributable to environmental factors. The table also provides a fine-grained analysis of duplicates and corrupted packages relative to the total number of transmitted packages. Disabling \ac{CRC} yields the lowest latency, followed by \SI{8}{\bit} \ac{CRC}, and then \SI{16}{\bit} \ac{CRC}. This aligns with our expectations, as fewer bits are sent and less time is spent on checksum computation and verification. However, disabling \ac{CRC} introduces data integrity issues. Without \ac{CRC}, 735 packets were received, including one duplicated package and 15 corrupted packets, resulting in a total success rate of $\tfrac{735 - 15 - 1}{750} \times 100 = 95.87\%$. For our target of wireless time synchronization between nodes, the primary requirement is a successful payload reception, as packets are used for low-latency signaling rather than data transmission. Consequently, occasional payload corruption is acceptable in this application and therefore \ac{CRC} is disabled for the \ac{OLCfg}.

\begin{table}
\centering
\caption{Summary of latency measurements across all evaluated ESB configurations.}
\label{tab:esb_latency_summary}
\begin{threeparttable}
\begin{tabularx}{\columnwidth}{l *{2}{>{\centering\arraybackslash}X}}
\toprule
\textbf{Configuration} & \textbf{Mean $\pm$ SD [$\mu$s]} & \textbf{Median [$\mu$s]} \\
\midrule
\multicolumn{3}{@{}l}{\textbf{CRC Mode}} \\
16-bits CRC  & 642.87 $\pm$ 221.35 & 562.39 \\
8-bits CRC   & 614.18 $\pm$ 166.12 & 558.39 \\
CRC off      & 602.43 $\pm$ 160.14 & 554.30 \\
\midrule
\multicolumn{3}{@{}l}{\textbf{Protocol Mode}} \\
dynamic      & 590.82 $\pm$ 111.39 & 562.39 \\
static       & 603.64 $\pm$ 146.58 & 563.10 \\
\midrule
\multicolumn{3}{@{}l}{\textbf{Bitrate Mode}} \\
2~Mbit/s (BLE) & 594.45 $\pm$ 156.24 & 559.32 \\
2~Mbit/s       & 607.45 $\pm$ 153.66 & 562.39 \\
\midrule
\multicolumn{3}{@{}l}{\textbf{TX Modes}} \\
automatic       & 570.21 $\pm$ 143.55 & 532.49 \\
manual          & 574.19 $\pm$ 149.56 & 534.23 \\
manual start    & 562.95 $\pm$ 128.19 & 534.23 \\
\midrule
\multicolumn{3}{@{}l}{\textbf{TX Power Levels}} \\
10 dBm      & 619.68 $\pm$ 202.84 & 548.67 \\
5 dBm       & 626.90 $\pm$ 204.34 & 549.79 \\
0 dBm       & 633.30 $\pm$ 212.35 & 552.25 \\
-12 dBm     & 598.16 $\pm$ 164.95 & 548.87 \\
-30 dBm     & 626.84 $\pm$ 212.04 & 550.66 \\
-70 dBm     & 621.74 $\pm$ 203.19 & 550.00 \\
\midrule
\multicolumn{3}{@{}l}{\textbf{Payload Construction}} \\
standard mode\tnote{\dag}   & 692.96 $\pm$ 186.54 & 625.16 \\
optimized mode\tnote{\ddag} & 683.43 $\pm$ 177.47 & 617.89 \\
\bottomrule
\end{tabularx}
\begin{tablenotes}[flushleft]
\footnotesize
\item All measurements are based on the timing difference between D0 and D7. For each config., 3–5 rounds were performed, yielding 450–750 data points.
\item[\dag] Payload constructed on \ac{APP-core} and transferred via \ac{IPC} to \ac{NET-core}.
\item[\ddag] Pre-constructed payload on \ac{NET-core} ready for immediate transmission.

\vspace{-5mm}

\end{tablenotes}
\end{threeparttable}
\end{table}

\subsubsection{Protocol Mode}
Table~\ref{tab:esb_latency_summary} shows a latency difference of \SI{8.85}{\micro\second} between protocol modes. While small, this difference is relevant for time-sensitive applications. Therefore, the dynamic mode is selected for the \ac{OLCfg}.

\subsubsection{Bitrate Mode}
Our analysis focuses on the two \qty{2}{\mega\bit\per\second} configurations, as this bitrate is significantly faster than \qty{1}{\mega\bit\per\second} and more relevant for latency optimization. Table~\ref{tab:esb_latency_summary} shows that both configurations perform similarly, with the \ac{BLE}-optimized mode achieving a slightly lower mean and median latency. This suggests that using \qty{2}{\mega\bit\per\second} with \ac{BLE} parameters is preferable for reducing overall latency.

\subsubsection{\ac{TX} Mode}
Table~\ref{tab:esb_latency_summary} shows that the manual start \ac{TX} mode achieves the lowest mean latency and \ac{SD}, making it the preferred choice for minimizing overall latency.

\subsubsection{\ac{TX} Power Levels}
Table~\ref{tab:esb_latency_summary} shows no clear trend in latency across \ac{TX} power levels, likely due to the short \qty{1}{\meter} test distance. \revise{To clarify the impact of this parameter, further evaluations over longer ranges are needed.
To estimate the upper bound in latency, a \ac{TX} power of \SI{0}{\dbm} was used as it produced the biggest median timing discrepancy.}

\subsection{Payload Construction}
While not an \ac{ESB} parameter, optimized payload generation can significantly reduce latency (see Section~\ref{sec:parameters}). For a fair comparison, both the standard and optimized modes used a one-byte payload. Table~\ref{tab:esb_latency_summary} shows that the optimized implementation achieves slightly lower latency than the standard mode.

\subsection{\acf{OLCfg}}
After evaluating all parameters, the configurations yielding the shortest transmission latencies were identified (see Table~\ref{tab:opt_latency_conf}). In addition, a one-byte payload was selected for testing the \ac{OLCfg}, both due to the use of the optimized payload construction mode and the reduced transmission delay associated with minimal payloads.
Table~\ref{tab:esb_latency_summary} shows that the \ac{OLCfg} achieves significantly lower latencies than any individual parameter variation tested.

\begin{table}
\centering
\caption{Selected configuration parameters for optimal latency performance.}
\label{tab:opt_latency_conf}
\begin{threeparttable}
\begin{tabularx}{\columnwidth}{l>{\centering\arraybackslash}X}
\toprule
\textbf{Parameter} & \textbf{Configuration} \\
\midrule
\ac{CRC} Mode        & disabled \\
Protocol Mode        & dynamic \\
Bitrate Mode         & \qty{2}{\mega\bit\per\second} \\
\ac{TX} Mode         & manual \\
\ac{TX} Output Power & \SI{0}{\dbm} \\
\midrule
Packet Construction Mode\tnote{\dag} & optimized mode \\
\bottomrule
\end{tabularx}
\begin{tablenotes}[flushleft]
\footnotesize
\item[\dag] Optimized mode minimizes header overhead compared to the default packet structure.

\end{tablenotes}
\end{threeparttable}
\end{table}

\begin{table}
\centering
\caption{Latency measurements for time differences across multiple input channels.}
\label{tab:olcnf}
\begin{threeparttable}
\begin{tabularx}{\columnwidth}{l *{2}{>{\centering\arraybackslash}X}}
\toprule
\textbf{Time Intervals} & \textbf{Mean $\pm$ \ac{SD} [\si{\micro\second}]} & \textbf{Median [\si{\micro\second}]} \\
\midrule
\textbf{\ac{D}0--\ac{D}7} & 504.99 $\pm$ 96.89 & 486.30 \\
\textbf{\ac{D}2--\ac{D}5} & 311.78 $\pm$ 96.90 & 293.07 \\
\textbf{\ac{D}3--\ac{D}4} & 204.27 $\pm$ 96.91 & 185.86 \\
\bottomrule

\vspace{-8mm}

\end{tabularx}
\end{threeparttable}
\end{table}

The transmission latency from the \ac{TX} \ac{APP-core} to the \ac{RX} \ac{APP-core} (D0 - D7) is \SI{504.99}{\micro\second}, while the latency from the \ac{TX} \ac{NET-core} to the \ac{RX} \ac{NET-core} (D2 - D5) is lower at \SI{311.78}{\micro\second}. The \ac{NET-core}-to-\ac{NET-core} path is already fully optimized, leaving limited further improvement potential.
In contrast, the higher \ac{APP-core} to \ac{APP-core} latency indicates that additional performance gains may be achievable through optimization of inter-core \ac{IPC} communication.

\section{Conclusion}\label{sec:conclusion}

\revise{This paper presents an accurate latency evaluation of the \ac{ESB} protocol for low-latency command broadcasting in multi-node systems,
achieving a consistent transmission latency, measuring \SI{311.78}{\micro\second} for \ac{NET-core}-to-\ac{NET-core} and \SI{504.99}{\micro\second} for \ac{APP-core}-to-\ac{APP-core} transfers.}
In contrast, \ac{BLE} shows higher and less predictable latency, as it depends on the connection interval (\SI{7.5}{\milli\second} minimum) between exchanges. Transmission latency can range from nearly zero to an entire interval, without application-layer control of timing.
However, with the \ac{ESB} protocol, the \ac{RX} must remain active and listening, trading deterministic latency for a significant decrease in energy efficiency.
Our results demonstrate that \ac{ESB} delivers deterministic, low-latency performance, making it well-suited for time-sensitive and performance-critical applications.
Hardware constraints of the \textsc{nRF5340} prevented the use of the fast ramp-up feature, which could further reduce the latency. 

\revise{Future work will investigate fast-ramp-up using a capable \textsc{nRF52} \ac{MCU}, while also considering effects from different acquisition setups, such as body shadowing and external device interference.}

\FloatBarrier

\bibliographystyle{IEEEtranDOI} 
\bibliography{references.bib}
\end{document}